\documentclass[prd,eqsecnum,twocolumn,showpacs,nofootinbib]{revtex4}

\usepackage{amsmath}
\usepackage{amssymb}
\usepackage{graphicx}

\newcommand{\be}{\begin{equation}}
\newcommand{\ee}{\end{equation}}

\newcommand{\msum}{\sum_{m=0}^\infty \!{}^{^\prime}}
\newcommand{\jsum}{\sum_{j=0}^\infty \!{}^{^\prime}}

\newcommand{\ppsum}{\sum_{m,j\in \mathbb{Z}}\!\!{}^{''}}

\newcommand{\zint}{\int_0^\infty}

\newcommand{\Ec}{\mathcal{E}}
\newcommand{\Et}{{\tilde{\mathcal{E}}}}

\newcommand{\Eh}{\hat{\mathcal{E}}}

\newcommand{\Emn}{\mathcal{E}_{m0}}
\newcommand{\Eb}{\bar{\mathcal{E}}}
\newcommand{\Ebmn}{\bar{\mathcal{E}}_{m0}}
\newcommand{\Einf}{\mathcal{E}_\infty}
\newcommand{\hT}{\hat{\mathcal{T}}}
\newcommand{\mK}{\mathcal{K}}
\newcommand{\mS}{\mathcal{S}}
\newcommand{\mH}{\mathcal{H}}
\newcommand{\emj}{{e}_{m,j}}
\newcommand{\temj}{\tilde{e}_{m,j}}

\newcommand{\tphi}{\tilde{\varphi}}
\newcommand{\ptau}{\partial_\tau}
\newcommand{\ptaus}{\partial^2_\tau}

\newcommand{\lsx}{\lambda^2_{mp}(x)}
\newcommand{\ls}{\lambda^2_{mp}}
\newcommand{\half}{{\textstyle\frac1{2}}}

\newcommand{\cyl}{\mathrm{cyl}}

\begin{document}

\title{Casimir effect at nonzero temperature for wedges and cylinders}

\date{\today}
\author{Simen {\AA}. \surname{Ellingsen}}
\email{simen.a.ellingsen@ntnu.no}
\author{Iver \surname{Brevik}}\email{iver.h.brevik@ntnu.no}
\affiliation{Department of Energy and Process Engineering, Norwegian
University of Science and Technology, N-7491 Trondheim, Norway}
\author{Kimball A. Milton}\email{milton@nhn.ou.edu}
\affiliation{Oklahoma Center for High Energy Physics and 
Homer L. Dodge Department of
Physics and Astronomy, The University of Oklahoma, Norman, OK 73019, USA}

\begin{abstract}
  We consider the Casimir-Helmholtz free energy at nonzero temperature $T$ for 
a circular cylinder and perfectly conducting wedge closed by a cylindrical arc,
either perfectly conducting or isorefractive. The energy expression at nonzero 
temperature may be regularized to obtain a finite value, except for a  
singular corner term in the case of the wedge which is present also at zero 
temperature. Assuming the medium in the interior of the cylinder or wedge be 
nondispersive with refractive index $n$, the temperature dependence enters 
only through the non-dimensional parameter $2\pi naT$, $a$ being the radius of 
the cylinder or cylindrical arc. We show explicitly that the known zero 
temperature result is regained in the limit $aT\to 0$ and that previously 
derived high temperature asymptotics for the cylindrical shell are reproduced 
exactly.
\end{abstract}

\maketitle

\section{Introduction}

The Casimir effect \cite{casimir48} is the name given to energies and forces 
due to field fluctuations in the presence of boundaries. Once a 
theoretical curiosity, the effect has gained enormous and still increasing 
attention since its first quantitative measurement a good decade ago
\cite{lamoreaux97}. Reviews of recent progress include 
Refs.~\cite{BookBordag09, milton04, BookMilton01}.

The first geometry, considered in Casimir's classic paper \cite{casimir48} 
was that of two perfectly conducting plates, generalized to arbitrary 
dispersive materials by Lifshitz \cite{lifshitz55}. The force between parallel 
plates of any purely dielectric material is attractive, and it was therefore 
surprising when it was shown by Boyer that the Casimir stress on a perfectly 
conducting spherical shell is repulsive \cite{boyer68}. 

While it was clear from Boyer's result that the Casimir effect has a strong 
geometry dependence, results for new geometries were slow in coming for a 
long time, and it was only in 1981 that DeRaad and Milton calculated the 
Casimir energy for a circularly cylindrical shell \cite{deraad81}. Since then 
a number of analytical efforts have added to the knowledge of Casimir effect 
in cylindrical cavities, both perfectly conducting \cite{gosdzinsky98,milton99,
lambiase99,razmi09} and (magneto)dielectric \cite{brevik94,nesterenko99,
caveroPelaez05,romeo05,caveroPelaez06,brevik07}.  Most treatments of the 
cylindrical geometry have dealt with the zero temperature situation, and only 
a few calculations have concerned finite temperature \cite{balian78, bordag02a,
bordag02b}, and in these references only the high-temperature asymptotics were 
derived. No analytical expression valid for all temperatures exists for the 
cylindrical geometry to our knowledge.

A related geometry is the wedge. First considered with respect to Casimir 
effect in the 1970s \cite{dowker78,deutsch79}, it has been the subject of 
several treatments later \cite{brevik96,brevik98,brevik01,nesterenko02,
razmi05}. The geometry is inviting in that it is analytically solvable and 
contains the geometries of parallel plates and a single semi-infinite plate as 
limiting cases. The geometry of a wedge intercut by a cylindrical shell was 
considered by Nesterenko and co-workers \cite{nesterenko01, nesterenko03} and 
energy densities in the same geometry were calculated by Saharian and 
co-workers \cite{rezaeian02,saharian07,saharian09}. We are not aware of any 
previous efforts to tackle the Casimir energy problem for a wedge at non-zero 
temperature. 

We recently revisited the latter geometry to calculate the energy, at zero 
temperature, of a perfectly conducting wedge closed by a cylindrical boundary, 
either perfectly conducting or magnetodielectric \cite{brevik09}. We showed how
that energy could be written on the form (subscript $0$ indicates zero 
temperature)
\be
  \Ec_0 = \Et_0(p) + \Eh,
\ee
where $\Et_0$ is a finite, regularizable energy closely analogous to that 
found for a cylinder \cite{deraad81, milton99}, whereas $\Eh$ is a divergent 
term associated with the corners where the arc meets the wedge.

Here and in the
following we will make frequent use of the symbol
\be
  p = \pi/\alpha,
\ee
where the physical range is $p\in[\half,\infty)$, but
which we will in general allow to take any real positive value. Throughout our 
calculations we set $c=\hbar=k_\mathrm{B}=1$.
It was shown \cite{brevik09} that $\Eh$ could be rendered finite provided the 
arc become transparent at high frequencies. 

The calculations in \cite{brevik09} were extended to the first consideration of
a wedge which is not perfectly conducting but instead assumed to be 
isorefractive (diaphanous), i.e., spatially uniform speed of light 
\cite{ellingsen09}. In that case the term $\Eh$ is not present at all. The 
diaphanous wedge is analogous to the system of an annular region between two 
perfectly conducting cylinders, intercut by two semitransparent, radially 
directed interfaces \cite{Milton:2009bz,Wagner:2009vq}.
 Notably, while the energy expressions for 
a perfectly conducting wedge or circularly cylindrical shell require some 
regularization scheme in order to give numerical meaning, the energy expression
obtained for the diaphanous wedge is immediately finite.   A review of 
the Casimir wedge problem and an early exposition of 
the issue we elaborate herein are found in Ref.~\cite{brevik10}.

Naturally, for the geometry of a perfectly conducting cylinder there is no 
divergent term $\Eh$ since there are no sharp corners. It turns out (c.f.\ 
the discussion in Section III of Ref.~\cite{brevik09}) that the Casimir energy 
of a perfectly conducting cylindrical shell is
\be\label{wedgecyl}
  \Et_\cyl = 2 \Et(p=1).
\ee
Thus, all of the calculations in the following sections, which are carried out 
for general $p$, are valid also for a cylindrical shell by letting $p\to 1$ 
and multiplying by an overall factor of $2$.

In the following we derive an analytical expression for the Casimir energy of 
a perfectly conducting wedge (modulo a singular term as encountered in the 
past) and a perfectly conducting cylindrical shell, valid for arbitrary opening
angles and all temperatures. This extends the calculations for the perfectly 
conducting wedge presented in \cite{brevik09}, and simultaneously those for a 
circularly cylindrical conducting shell \cite{deraad81, milton99}, to the case 
of finite temperature. We show how the energy expression, which for $T>0$ is 
the Helmholtz free energy, may be regularized by a scheme of Epstein-zeta 
functions to obtain a numerically useful expression. We show explicitly that 
the expression thus obtained reduces to the previously derived zero temperature
limit, and that the two leading terms of the high temperature asymptotic 
expansion, derived by Bordag, Nesterenko and Pirozhenko \cite{bordag02b}, are 
reproduced exactly as a special case.

\section{Casimir-Helmholtz Free Energy of Wedge and Cylinder}

\begin{figure}[tb]
  \includegraphics[width=3.4in]{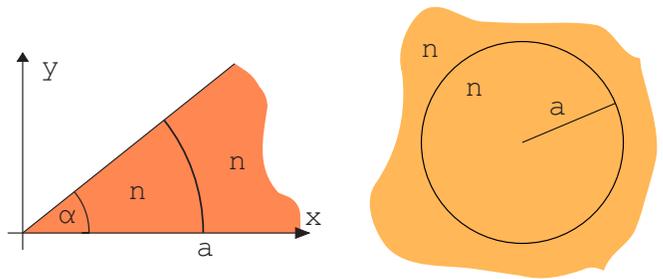}
  \caption{The geometry considered: (left) a wedge of opening angle 
$\alpha$ closed by a cylindrical shell at radius $a$. The results are 
automatically applicable to a cylindrical shell of radius $a$ (right) when 
$\alpha=\pi$ [equation (\ref{wedgecyl})].}
  \label{fig_geom}
\end{figure}

We take as our starting point the zero temperature energy derived for the 
geometry of a perfectly conducting wedge of opening angle $\alpha$ closed by a 
perfectly conducting cylindrical arc of radius $a$, derived in \cite{brevik09},
shown on the left side of Fig.~\ref{fig_geom}. 

Henceforth we shall focus on the term $\Et_0$, which may be written 
\cite{brevik09}
\be
  \Et_0 = \msum \Emn
\ee
with
\begin{align}
  \Emn =& -\frac1{8\pi^2}\int_{-\infty}^\infty dk \int_{-\infty}^\infty d\zeta 
\zeta\notag \\
  &\times\frac{d}{d\zeta} \ln I_{mp}(x)I'_{mp}(x)K_{mp}(x)K'_{mp}(x)
\end{align}
where we define the shorthand $x^2 = a^2(k^2 + n^2 \zeta^2)$ where $n$ is the 
index of refraction of the medium inside the wedge. We assume $n$ to be 
constant with respect to $\zeta$ and uniform in space. Here $\omega = i\zeta$ 
is the reciprocal of imaginary (Euclidian) time.
By means of partial integration with respect to $\zeta$, adding a trivial 
constant and noting that the integrand is symmetrical under $\zeta \to -\zeta$ 
and $k\to -k$, this may be written on the familiar form
\be\label{Ez}
  \Emn = \frac1{2\pi^2}\zint dk \zint d\zeta \ln[1-x^2\lsx],
\ee
wherein we use the shorthand
\be
  \lambda_{\nu}(x) = \frac{d}{dx}\left[I_\nu(x)K_\nu(x)\right].
\ee

The Helmholtz free energy at $T>0$ is obtained from the `trace-log' formula 
(\ref{Ez}) by compactifying the Euclidean time axis as is well known. 
Technically this amounts to the transition
\be\label{finiteT}
  \zint d\zeta f(\zeta) \to 2\pi T \jsum f(\zeta_j),
\ee
where $\zeta_j = 2\pi j T$ are the Matsubara frequencies. Changing the
integration variable from axial momentum $k$ to $x$, the resulting expression 
for the finite part of the free energy may be written
\be
  \Et = \frac{T}{\pi a}\msum\jsum\emj,
\ee
where
\be\label{emj}
\emj(\tau,p) = \int_{j\tau}^\infty \frac{dx\, x}{\sqrt{x^2-j^2\tau^2}}
\ln[1-x^2\lsx],
\ee
and where we have defined the dimensionless temperature 
\be
  \tau = 2 \pi n a T.
\ee
Similarly to the case at zero temperature this simple expression is in need of 
regularization in order to give numerical meaning since it is formally 
divergent. 

\section{Regularization of the free energy expression}

We here follow a scheme closely reminiscent of that of Milton and DeRaad 
\cite{deraad81}, and particularly Milton, Nesterenko and Nesterenko 
\cite{milton99} (cf.\ also Appendix A of Ref.~\cite{brevik09}). 

As follows from the uniform asymptotic expansion of modified cylindrical 
Bessel functions, e.g.\ \S 9.7 of Ref.~\cite{BookAbramowitz64}, 
the logarithmic 
factor in the integrand of (\ref{emj}) has the asymptotic behavior
\begin{subequations}\label{lnexp}
\begin{align}
  \ln[1-x^2 \ls] &\sim -\frac{x^4}{4(m^2 p^2 + x^2)^3}, &m,x\to \infty;
\label{masymp} \\
  \ln[1-x^2 \lambda_0^2]&\sim -\frac{x^4}{4(1 + x^2)^3}, &x\to \infty.
\end{align}
\end{subequations}

To see how this behavior gives rise to a formal divergence, consider the case 
of large $m$ for which
\begin{subequations}
\begin{align}
\emj \sim& -\frac1{4}\int_{j\tau}^\infty \frac{dx \, x^5}{\sqrt{x^2-j^2
\tau^2}(m^2p^2 + x^2)^3}\notag \\
=& -\frac1{4}\zint \frac{dy (y^4 + 2 j^2\tau^2y^2 + j^4\tau^4)}{(m^2p^2 + j^2
\tau^2 + y^2)^3}\label{yasymp} \\
=& -\frac{3\pi}{64 \varphi} -\frac{\pi j^2\tau^2}{32 \varphi^3} 
-\frac{3\pi j^4\tau^4}{64 \varphi^5}.\label{asymTerms}
\end{align}
\end{subequations}
where we substituted $y^2 = x^2 - j^2\tau^2$ and defined the shorthand 
\be\label{phi}
  \varphi = \sqrt{p^2 m^2 + \tau^2 j^2}.
\ee
The three terms of (\ref{asymTerms}) correspond to the three terms of the 
integrand of (\ref{yasymp}).
All of the terms of (\ref{asymTerms}) clearly diverge when summed over $j$ and 
$m$. 

The first step in regularization is to add and subtract the asymptotic 
behavior (\ref{lnexp}) in the form (\ref{yasymp})
\be
  \Et =\Eb + \Einf\label{addsubtract}
\ee
where we define the energy with the leading asymptotic term subtracted,
\begin{subequations}\label{ebdef}
\begin{align}
  \Eb =& \frac{T}{\pi a}\msum\jsum\temj \label{Ebar}\\
\temj =& \int_{j\tau}^\infty \frac{dx\, x}{\sqrt{x^2-j^2\tau^2}}  \notag \\
  &\times\left\{\ln[1-x^2\lsx]+ \frac{x^4}{4(m^2p^2 + x^2)^3}\right\}; \\
\tilde{e}_{0,j} =& \int_{j\tau}^\infty \frac{dx\, x}{\sqrt{x^2-j^2\tau^2}}  
\notag \\
  &\times\left\{\ln[1-x^2\lambda_0^2(x)]+ \frac{x^4}{4(1 + x^2)^3}\right\},
\end{align}
\end{subequations}
and the additional, non-regularized energy
\be
\Einf =- \frac{T}{4\pi a}\msum\jsum\zint\frac{dy(y^{4} + 2 j^2\tau^2y^{2} 
+ j^4\tau^4)}{(\tphi^2 + y^2)^3} \label{Einfinity}.
\ee
wherein
\be
\tphi = \left\{\begin{array}{cc}\sqrt{1 + \tau^2 j^2}, &m=0 \\ \sqrt{p^2 m^2 
+ \tau^2 j^2}, & m\geq 1\end{array} \right. .
\ee

To regularize $\Einf$ we introduce the small quantity $s$ and write
\begin{align}
\Einf =&- \frac{T}{4\pi a}\lim_{s\to 0^+}\msum\jsum\zint\frac{dyy^{-s}}
{(\tphi^2 + y^2)^3}\notag\\
  &\times (y^{4} + 2 j^2\tau^2y^{2} + j^4\tau^4) \notag \\
\sim& -\frac{T}{64 a}\lim_{s\to 0^+}\msum\jsum\left\{\frac{3}{\tphi^{1+s}}
\right.\notag \\
&\left.+\frac{2j^2\tau^2}{\tphi^{3+s}}+\frac{3j^4\tau^4}{\tphi^{5+s}}\right\},
\end{align}
where as in Eq.~(\ref{asymTerms}) we have used the evaluation,
\begin{equation}
  \zint \frac{dy y^{4-s}}{(\tphi^2 + y^2)^3} = \frac{\pi(1-s)(3-s)}
{16 \tphi^{1+s}}\sec\frac{\pi s}{2},
\end{equation}
which is valid for $-1<s<5$, so it may be used for $s$ near 0, near 2, or
near 4.
We use the relations [$\partial_\tau = \partial/\partial \tau$]
\begin{subequations}
\begin{align}
  \partial_\tau \frac1{\tphi^q} =& -\frac{q\tau j^2}{\tphi^{q+2}};\\
\partial^2_\tau \frac1{\tphi^q} =& -\frac{q j^2}{\tphi^{q+2}} + \frac{q (q+2) 
\tau^2j^4}{\tphi^{q+4}}
\end{align}
\end{subequations}
to write
\begin{align}\label{diffEinf}
\Einf =&-\frac{T}{64 a}(3-3\tau\ptau + \tau^2\ptaus)\notag \\
&\times\lim_{s\to 0^+} \msum\jsum\frac1{\tphi^{1+s}}.
\end{align}

The sum in (\ref{diffEinf}) can be regularized by analytical continuation. 
We will write it in the following form, using symmetry properties with respect 
to $m\leftrightarrow -m$ and $j\leftrightarrow -j$:
\begin{align}
\lim_{s\to 0^+} \msum\jsum\frac1{\tphi^{1+s}} =& \frac1{4} \lim_{s\to 0^+}
\sum_{m,j\in \mathbb{Z}}\frac1{\tphi^{1+s}}\notag \\
  =& \frac1{4} + \frac1{4} \mS(\tau,p) + \frac1{2}\mK(\tau).
\end{align}
Here we have defined
\be\label{sdef}
 \mS(\tau,p) =\lim_{s\to 0^+}\ppsum\frac1{\varphi^{1+s}}=\mS(p,\tau),
\ee
wherein the double prime on the summation mark means that the term $m=j=0$ is 
explicitly excluded, and 
\begin{align}
\mK(\tau) =& \lim_{s\to 0^+}\sum_{j=1}^\infty\left[\frac{1}{\tphi^{1+s}}
-\frac{1}{\varphi^{1+s}}\right]_{m=0}\notag \\
=& \sum_{j=1}^\infty \frac{j\tau - \sqrt{1+j^2\tau^2}}{j\tau
\sqrt{1+j^2\tau^2}}.\label{kform}
\end{align}
Clearly, $\mK(\tau)$ is finite for all $\tau> 0$. 

The function $\mS(\tau,p)$ may be regularized by use of the Chowla-Selberg 
formula [see e.g.\ Eq.\ (4.33) of Ref.~\cite{elizalde95}]
\begin{align}
  \ppsum &(a m^2 + bmj + cj^2)^{-q} = 2\zeta(2q)a^{-q}\notag \\
  &+ \frac{2^{2q}\sqrt{\pi}a^{q-1} \Gamma(q-\half)\zeta(2q-1)}{\Gamma(q)
\Delta^{q-\frac1{2}}}\notag\\
  &+\frac{2^{q+\frac{5}{2}} \pi^q}{\Gamma(q)\Delta^{\frac{1}{2}
(q-\frac{1}{2})}\sqrt{a}}\sum_{l=1}^\infty l^{q-\frac{1}{2}} 
\sigma_{1-2q}(l)\notag \\
  &\times \cos(l\pi b/a) K_{q-\frac{1}{2}}(\pi l \sqrt{\Delta}/a),
\label{chowlaseberg}
\end{align}
where
\begin{align}
  \Delta =& 4 a c - b^2, \\
  \sigma_w(l) =& \sum_{\nu|l}\nu^w,
\end{align}
where $\nu$ are summed over the divisors of $l$ and it is assumed that 
$\Delta>0$. $K$ is again the modified Bessel function of the second kind. The 
apparent pole as $q\to \half$ now vanishes due to a cancellation between the 
first two terms of (\ref{chowlaseberg}), and we find that letting $q=\half + 
{\textstyle \frac{s}{2}}$ and taking the limit $s\to 0^+$ (here $a=p^2, b=0, 
c=\tau^2$)
\be\label{sform}
  \mS(\tau,p) = \frac{2}{p}(\gamma - \ln\frac{4\pi p}{\tau})+\frac{8}{p}
\sum_{l=1}^\infty \sigma_0(l)K_0(2\pi l \tau/p)
\ee
where $\gamma = 0.577216...$ is Euler's constant. 
Now $\sigma_0(l)$ is simply the number of positive divisors of $l$, 
$\sigma_0(1)=1, \sigma_0(2)=\sigma_0(3)=2, \sigma_0(4)=3$ etc.
Note that Eq.~(\ref{sform}) is valid for all $\tau$; although it appears
most convenient for large $\tau$, it is, by the symmetry property seen
in Eq.~(\ref{sdef}), equally useful for small $\tau$.

We finally write down the final, regularized energy of the wedge (and, 
simultaneously, cylinder) at finite $T$, using
the convention used in Ref.~\cite{brevik09} 
\be
  \Et(\tau,p,a) = \frac{1}{8\pi n a^2} e(\tau,p),
\ee
in terms of
\begin{align}
  e&(\tau,p) = \frac{4\tau}{\pi}\msum\jsum \temj(\tau,p) \notag \\
&- \frac{\tau}{64}(3-3\tau\ptau + \tau^2\ptaus)[1+ 2\mK(\tau) + \mS(\tau,p)].
\label{etaup}
\end{align}
with $\temj, \mK$, and $\mS$ given in Eqs.~(\ref{ebdef}), (\ref{kform}) and 
(\ref{sform}), respectively. The differentiations with respect to $\tau$ are 
now straightforward, should the full expanded expression be desirable.

\begin{figure}[tb]
  \includegraphics[width=2.4in]{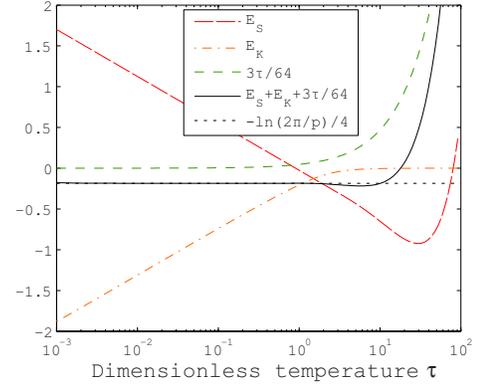}
  \caption{The additional terms of the regularized energy in 
Eq.~(\ref{etaup}) which are subtracted from the
double sum there in the case $p=3$.  Shown also is the sum of
the three additional terms and their low-temperature
asymptotic value from Eq.~(\ref{zerotaulim}). }
  \label{fig:additionterms}
\end{figure}

\begin{figure}[tb]
  \includegraphics[width=2.8in]{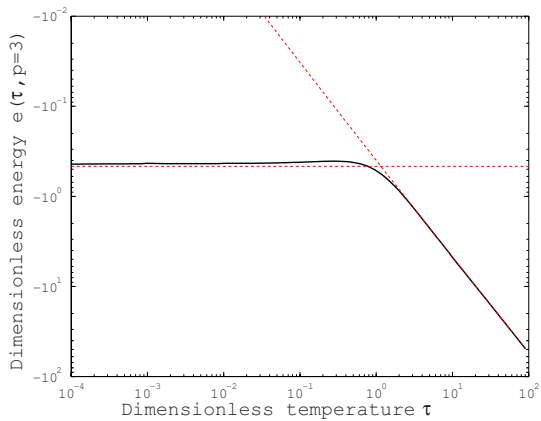}
  \caption{Dimensionless energy $e(\tau,p)$ for the case $p=3$, i.e.\ opening 
angle $\alpha=\pi/3$ approximated by a ``brute force'' calculation truncating 
the sums. The zero-temperature limit and high-$\tau$ asymptote are shown as 
dashed lines. }
  \label{fig:etau3}
\end{figure}

In Fig.~\ref{fig:additionterms} we plot the three additional terms in the 
second line of Eq.~(\ref{etaup}) where we have defined the shorthand
\begin{subequations}
\begin{align}
  \hT=&  (3-3\tau\ptau + \tau^2\ptaus);\\
  E_S(\tau,p) =& \frac{\tau}{64}\hT \mS(\tau,p);~~ E_K(\tau,p) = 
\frac{\tau}{32}\hT \mK(\tau).
\end{align}
\end{subequations}

Figure \ref{fig:etau3} shows a numerical calculation of $e(\tau,p=3)$ as a 
function of $\tau$ along with its high and low $\tau$ asymptotes (see 
derivations in the following sections). The calculation was performed by 
``brute force'' by truncating the sums after a number of terms, and has 
somewhat limited accuracy due to the large number of terms in the $j$ sum in 
Eq.~(\ref{etaup}) required for small $\tau$, scaling as $\tau^{-1}$.

\section{Regaining the limit of zero temperature}

Comparing Eq.~(\ref{etaup}) with the zero temperature result derived in 
Ref.~\cite{brevik09}, and previously known for the cylindrical shell 
\cite{deraad81, milton99}, it is not obvious that our expression simplifies to 
the zero temperature result as $\tau\to 0$. In this section we show that upon 
careful examination the correct limit is in fact obtained.

Let us write down the zero temperature result $\Et_0$ for general $p$ in its 
regularized form suitable for comparison\footnote{These definitions of $\Eb_0$ 
and $\Ebmn$ differ from those of Ref.~\cite{brevik09} by a prefactor $n$.} 
(c.f.\ Ref.~\cite{brevik09}, Eq.~(4.14)):
\begin{subequations}\label{zerotempReg}
\begin{align}
  \Et_0 =& \half\Eb_0 + \sum_{m=1}^\infty \Eb_m + 
\frac1{32\pi na^2}\ln(2\pi/p);\label{zerotemp}\\
  \Eb_0 =& \frac{1}{4\pi na^2}\zint dxx\left[\ln(1-x^2\lambda_0^2)
+\frac{x^4}{4(1+x^2)^3}\right]; \\
  \Ebmn =& \frac{1}{4\pi na^2}\zint dxx\left[\ln(1-x^2\lambda_{mp}^2)
+\frac{x^4}{4(m^2p^2+x^2)^3}\right].
\end{align}
\end{subequations}

The finite temperature quantity $\Eb$ of Eq.~(\ref{ebdef}) is analytic as 
$\tau\to 0$ and inverse application of the transition (\ref{finiteT}) simply 
gives us
\be
  \Eb \buildrel{\tau\to 0}\over{\longrightarrow} \half\Eb_0 
+ \sum_{m=1}^\infty \Eb_m.
\ee
What remains is essentially to determine the low $\tau$ behavior of 
$\mK(\tau)$ and $\mS(\tau,a)$ to check that the last term of 
Eq.~(\ref{zerotemp}) may be regained.

To study the behavior of $\mS$ it is convenient to employ the symmetry 
relation $\mS(\tau,p)=\mS(p,\tau)$ which gives
\be
  \mS(\tau, p)  = \frac{2}{\tau}(\gamma - \ln\frac{4\pi \tau}{p})
+\frac{8}{\tau}\sum_{l=1}^\infty \sigma_0(l)K_0(2\pi l p/\tau).\label{Staup}
\ee
For large arguments $K_0(x)\propto \exp(-x)$, so the sum over $l$ is 
exponentially small as $\tau\to 0$. This immediately gives the asymptotic 
behavior:
\be\label{Slotau}
  \mS(\tau, p)  \sim \frac{2}{\tau}[\gamma - \ln(2\pi/p)-\ln 2 - \ln\tau], 
~\tau\to 0.
\ee

Next we turn to $\mK(\tau)$. Using the Euler-Maclaurin formula (e.g.\ 
Ref.~\cite{BookAbramowitz64}, p.806) we have
\begin{align}
  \mK(\tau) =& \sum_{l=0}^\infty \varkappa(\tau + l\tau)= \frac1{\tau}
\int_\tau^\infty dt \varkappa(t) \notag \\
  & + \half \varkappa(\tau) - \sum_{l=1}^\infty \frac{\tau^{2l-1}B_{2l}}{(2l)!}
\varkappa^{(2l-1)}(\tau)\label{eulermaclaurin}
\end{align}
with
\be
  \varkappa(t) = \frac{t-\sqrt{1+t^2}}{t\sqrt{1+t^2}}.
\ee

The integral has the solution
\begin{align}
  \int_\tau^\infty dt \varkappa(t) =& \ln\frac{2\tau}{\tau 
+ \sqrt{1+\tau^2}}\notag \\
  =& \ln\tau + \ln 2-\tau+{\textstyle\frac1{6}}\tau^3+...
\end{align}
as $\tau\to 0$. Moreover, $\varkappa(\tau)=-1/\tau + 1-\half \tau^2+..$, and 
upon inspection we recognize that 
\begin{align}
  \tau^{2l-1}\varkappa^{(2l-1)}(\tau) =& \frac{(2l-1)!}{\tau}
  +(-1)^l\left[\frac{(2l)!}{2^l l!}\right]^2 \tau^{2l}+...
\end{align}
To leading order in $\tau$, thus, the sum in Eq.~(\ref{eulermaclaurin}) reads 
${\textstyle\frac1{\tau}}\sum_{l=1}^\infty B_{2l}/2l$. As is typically the case
for series expansions close to non-analytical points, the series is formally 
divergent. It can, however, be regularized by means of Borel summation 
\cite{BookBender99}. For a highly similar problem and details on 
how to approach it, see \cite{ellingsen08}. We show in Appendix A that the 
Borel regularized sum evaluates to
\be\label{borel}
  \sum_{l=1}^\infty \frac{B_{2l}}{2l} = \gamma - \half.
\ee
Thus we have found the low-$\tau$ expansion of $\mK(\tau)$:
\be\label{Klotau}
  \mK(\tau) \sim \frac1{\tau}(\ln\tau + \ln 2 - \gamma)-\half + \dots, 
~~\tau\to 0.
\ee
Further terms cancel at least to order $\tau^2$, the leading order correction 
being at least of order $\tau^4$. 

Combining (\ref{Slotau}) and (\ref{Klotau}) we find, to leading order in 
$\tau$, the expression in square brackets in (\ref{etaup}):
\be
  [1+ 2\mK + \mS] \sim  -\frac{2}{\tau}\ln(2\pi / p)+\mathcal{O}(\tau^4), 
~~\tau \to 0\label{zerotaulim}
\ee
and using 
\be
  (3-3\tau\ptau + \tau^2\ptaus)\frac1{\tau} = \frac{8}{\tau}
\ee
we regain exactly the zero temperature result (\ref{zerotempReg}). As 
illustrated in Fig.~\ref{fig:etau3} this limit is reached very rapidly 
as $\tau\to 0$. While we have ascertained in the above that the correction term
in (\ref{zerotaulim}) is at least of order $\tau^4$, there is reason to suspect
that the behavior is in fact exponential, as is the case for $\mS$ as seen 
from Eq.~(\ref{Staup}).


\section{High-$\tau$ asymptotics: agreement with previous results for 
cylindrical shell}\label{comparison}

We will finally
determine the asymptotic behavior in the limit $\tau\gg 1$. Here the 
contribution from $\Eb$ is given by the zeroth Matsubara term only. Consider 
the reduced energy $e(\tau,p)$ of Eq.~(\ref{etaup}) in which
\be
  \frac{4\tau}{\pi}\msum\jsum\temj(\tau,p) \sim \frac{2\tau}{\pi}C(p),
~~\tau\to \infty,
\ee
with
\begin{align}
  C(p) = \half\zint dx\left[\ln(1-x^2\lambda_0^2)+\frac{x^4}{4(1+x^2)^3}\right]
\notag \\
+ \sum_{m=1}^\infty\zint dx\left[\ln(1-x^2\lambda_{mp}^2)+\frac{x^4}
{4(m^2p^2+x^2)^3}\right].\label{Cp}
\end{align}
Some numerical values are
\begin{subequations}
\begin{align}
  C(1) =& -0.75814;\label{C1}\\
  C(2) =& -0.76558;\\
  C(3) =& -0.76645.
\end{align}
\end{subequations}
These values were obtained with Mathematica, including 100 terms in the sum 
while checking convergence.

The high-$\tau$ behavior of $\mS$ is given immediately by Eq.~(\ref{sform}):
\be
  \mS(\tau,p) \sim \frac{2}{p}(\gamma - \ln 4\pi p+\ln\tau),~~\tau\to \infty
\ee
where the correction term is exponential, wherewith
\[
  \tau\partial_\tau \mS\sim \frac{2}{p};~~ \tau^2\partial^2_\tau \mS\sim -\frac{2}{p}.
\]

To study the behavior of $\mK(\tau)$ and its derivatives it is useful to 
define $\beta=1/\tau$ and write $\mK(\tau) =\beta \mH(\beta)$ with
\be
\mH(\beta) = \sum_{j=1}^\infty \frac{j-\sqrt{j^2+\beta^2}}
{j\sqrt{j^2+\beta^2}}.
\ee
With a litte calculation one ascertains that
\be
  \tau(3-3\tau\partial_\tau+\tau^2\partial^2_\tau)\mK(\tau)=
(8+7\beta\partial_\beta+\beta^2\partial^2_\beta)\mH(\beta).
\ee
When $\beta\to 0$ it is simple to see from 
\begin{subequations}
\begin{align}
  \mH'(\beta) =& -\beta\sum_{j=1}^\infty (j^2+\beta^2)^{-\frac{3}{2}};\\
\mH''(\beta)=& \sum_{j=1}^\infty\frac{2\beta^2-j^2}{(j^2+\beta^2)^\frac{5}{2}};
\end{align}
\end{subequations}
that
\[
  \mH(\beta)\sim\half\beta\mH'(\beta)\sim\half\beta^2\mH''(\beta)\sim
-\half\beta^2\zeta(3),~~\beta\to 0.
\]
Hence we can safely ignore the term involving $\mK$ at high $\tau$.

Combining this, the high-$\tau$ behavior of $e(\tau,p)$ is
\begin{align}
  e(\tau,p) \sim& \tau\left[\frac{2C(p)}{\pi}-\frac{3p+6(\gamma-\ln4\pi p)-8}
{64p}\right.\notag\\
  &-\left.\frac{3}{32p}\ln\tau\right].\label{highT}
\end{align}

The high temperature asymptotics of perfectly conducting spherical and 
cylindrical shells with vacuum inside and outside were calculated by Bordag, 
Nesterenko and Pirozhenko \cite{bordag02a,bordag02b} using the method of heat 
kernel coefficients. They, like us, found that the two leading order terms were
of order $T$ and $T\ln T$ as $T\to \infty$. The latter of these terms had been 
worked out some time previously by Balian and Duplantier\cite{balian78}, who 
also found an approximate (though not very accurate) value for the former.

The result of the calculations reported in \cite{bordag02b} was, in our 
notation
\begin{align}
  \Et_\cyl\sim& -0.22924 \frac{T}{a} - \frac{3T}{64a}\ln \frac{aT}{2} 
+ \mathcal{O}(T^{-1})\notag\\
  =& -\frac{\tau}{8\pi a^2}\left[0.44237 + \frac{3}{16}\ln\tau + 
\mathcal{O}(\tau^{-2})\right].
\end{align}

As previously mentioned, $\Et_\cyl = 2\Et_{p=1}$. With the expansion 
(\ref{highT}) we find, using (\ref{C1}),
\begin{align}
  2\Et_{p=1} \sim& -\frac{\tau}{8\pi a^2}\left[\frac{6\gamma-5-6\ln 4\pi}{32}
-\frac{4}{\pi}C(1)\right.\notag\\
  &+\left.\frac{3}{16}\ln\tau+...\right]\notag \\
  =&-\frac{\tau}{8\pi a^2}\left[0.44270+\frac{3}{16}\ln\tau+...\right].
\label{ushit}
\end{align}
The slight numerical difference we believe to be due to the approximate 
numerical method used in \cite{bordag02b}. We show analytically in Appendix 
\ref{app:highT} that the correspondence is in fact exact.

\section{Wedge with diaphanous arc}

The above results can easily be extended to the case where the perfectly 
conducting arc is replaced by a diaphanous arc, that is, a magnetodielectric 
interface so that the product $n^2 = \varepsilon \mu$ is the same for radii 
both smaller than and greater than $a$. This geometry was considered at zero 
temperature in Ref.~\cite{brevik09}. The electromagnetic boundary conditions at
the arc separate in a simple way in this case and the dependence on material 
properties enter only through the reflection coefficient
\be
  \xi = \frac{\varepsilon_2-\varepsilon_1}{\varepsilon_2+\varepsilon_1}=
-\frac{\mu_2-\mu_1}{\mu_2+\mu_1}.
\ee

\begin{figure}[tb]
  \includegraphics[width=3.4in]{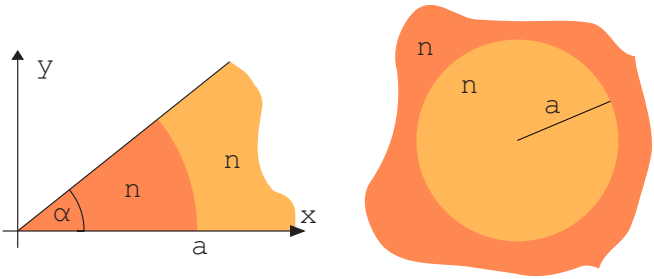}
  \caption{Same geometry as in figure \ref{fig_geom} but now with diaphanous 
instead of perfectly conducting arc, i.e., so that $n^2=\varepsilon\mu$ is the 
same both sides of the interface. We still assume nondispersive media.}
  \label{fig_diaph}
\end{figure}

The change in geometry leaves the energy expression (\ref{Ez}) unaltered but 
for the simple replacement
\be
  \ln[1-x^2\lsx]\to \ln[1-\xi^2x^2\lsx].
\ee
This merely introduces a prefactor $\xi^2$ in all correction terms, and we can 
write down the result for the diaphanous wedge, and simultaneously cylinder 
(by letting $p=1$ and multiplying by $2$ as discussed above), as:
\begin{subequations}
\begin{align}
  \Et_\xi(\tau,p,a) =& \frac{1}{8\pi n a^2} e_\xi(\tau,p);\\
  e_\xi(\tau,p) =& \frac{4\tau}{\pi}\msum\jsum \temj(\tau,p,\xi)- 
\frac{\tau\xi^2}{64}(3-3\tau\ptau \notag \\
& + \tau^2\ptaus)[1+ 2\mK(\tau) + \mS(\tau,p)],
\end{align}
wherein
\begin{align}
\temj(\tau,p,\xi) =& \int_{j\tau}^\infty \frac{dx\, x}{\sqrt{x^2-j^2\tau^2}}
\left\{\ln[1-\xi^2x^2\lsx]\right.  \notag \\
  &\left.+ \frac{\xi^2x^4}{4(m^2p^2 + x^2)^3}\right\};\\
\tilde{e}_{0,j}(\tau,\xi) =& \int_{j\tau}^\infty \frac{dx\, x}{\sqrt{x^2-j^2
\tau^2}}\left\{\ln[1-\xi^2x^2\lambda_0^2(x)]\right.  
\notag \\
  &\left.+ \frac{\xi^2x^4}{4(1 + x^2)^3}\right\}.
\end{align}
\end{subequations}

Since $\xi$ enters the correction terms from renormalization only through the 
prefactor, generalization of the weak-coupling expansions (to leading order in 
$\xi^2$) considered in \cite{brevik09} to nonzero $\tau$ is trivial.

\section{Concluding remarks}
We have given for the first time results for the temperature dependence of
the Casimir energy for a wedge, closed by a circular arc, all boundaries
being perfectly conducting.
This includes, as a special case, the perfectly conducting cylindrical shell
case.  (Except for that case, there is a divergent term, due to the corner
where the circular arc meets the wedge boundaries, which we here simply
omit.)  The low temperature result agrees with the zero-temperature result
found previously, except for what is probably an exponentially small
correction, while the high temperature result agrees with that of Bordag, 
Nesterenko, and Pirozhenko for the case of a cylinder \cite{bordag02b}.

\acknowledgments
The work of KAM was supported in part by grants from the US National
Science Foundation and the US Department of Energy.

\appendix

\section{Evaluation of Eq.~(\ref{borel}) by Borel summation}

To evaluate a (possibly divergent) series $ Z=  \sum_{l=1}^\infty a_l$ by 
Borel summation \cite{BookBender99} we define the function
\be
  \phi(x) = \sum_{l=1}^\infty\frac{a_l}{l!}x^l.
\ee
If $\phi(x)$ is finite for sufficiently small $x$, we define the Borel 
transform as
\be
  \mathcal{B}(x) = \zint dt e^{-t} \phi(xt),
\ee
from which the Borel regularized value of the sum $Z$ is $Z = \mathcal{B}(1)$. 
We consider the sum
\be\label{thesum}
\sum_{l=1}^\infty \frac{B_{2l}}{2l} = \frac1{2} + \sum_{l=1}^\infty 
\frac{B_l}{l},
\ee
since $B_1=-1/2$ and $B_3=B_5=B_7=...=0$. The Borel transform
 of the latter sum is thus
\be
\mathcal{B}(x) = \zint dt e^{-t} \sum_{l=1}^\infty \frac{B_l}{l\cdot l!}(xt)^l.
\ee
The generating function of the Bernoulli numbers is
\be
  \sum_{l=1}^\infty \frac{B_l}{l!}y^l = \frac{y}{e^y-1}-1
\ee
to evaluate
\begin{align}
  \frac{d\mathcal{B}}{dx}(x) =& \frac1{x}\zint dt e^{-t} \sum_{l=1}^\infty 
\frac{B_l}{l!}(xt)^l=\zint \frac{dt\, te^{-t}}{e^{xt}-1}-\frac1x\notag \\
  =& x^{-2}\zint \frac{duu e^{-u \frac{x+1}{x}}}{1-e^{-u}}-\frac1x
 = x^{-2}\psi^{(1)}({\textstyle \frac{x+1}{x}})-\frac1x,
\end{align}
where $\psi^{(n)}(x)$ is the polygamma function, whose integral representation 
was recognized (Ref.~\cite{BookAbramowitz64} Eq.~6.4.1) by making the 
substitution $u=xt$. Thus we evaluate the integral to
\begin{align}
\mathcal{B}(x)+\ln x 
=& \int^x \frac{dy}{y^{2}}\psi^{(1)}({\textstyle \frac{y+1}{y}})
 = -\int^{\frac{x+1}{x}} dv \,\psi^{(1)}(v) \notag\\
  =& -\psi({\textstyle \frac{x+1}{x}})+\mbox{constant},
\end{align}
where from the requirement that $\mathcal{B}(0)=0$ we see that the
integration constant is zero.
Thus we find the Borel value of the sum (\ref{thesum}) to be
\be
\sum_{l=1}^\infty \frac{B_{2l}}{2l} = \frac1{2} + \mathcal{B}(1) =\gamma-\half,
\ee
noting that $\psi(2)=1-\gamma$.

\section{Correspondence with high-$T$ asymptotics for the cylinder in vacuum}
\label{app:highT}

The heat kernel expansion for high temperatures calculated in 
Ref.~\cite{bordag02b} for the cylindrical shell in vacuum begins
\begin{align}
\Et_\cyl \sim& -\frac{T}{2}\zeta'(0) - \frac{a_{3/2}}{(4\pi)^{3/2}}T\ln T + 
\dots\notag \\
=& -\frac{\tau}{8\pi a^2}\left[2a\zeta'(0)-\frac{3}{16}\ln2\pi a + \frac{3}{16}
\ln\tau+...\right]
\end{align}
where the `zeta determinant' $\zeta'(0)$ is a constant defined in Ref.~\cite{bordag02b} and we have inserted their value \cite{bordag02a,bordag02b}
\be
  \frac{a_{3/2}}{(4\pi)^{3/2}} = \frac{3}{64a}.
\ee
The term proportional to $\tau\ln\tau$ is obviously identical to our 
expression in Eq.~(\ref{ushit}). We consider only the term linear in 
$\tau$. Comparison with (\ref{ushit}) gives, with minimal manipulation, that 
the asymptotes correspond exactly according to $\Et_\cyl=2\Et_{p=1}$, provided
\be
  \pi a \zeta'(0)+2C(1) = \frac{\pi}{64}[6\gamma - 5 + 6 \ln \frac{a}{2}].
\ee
In Appendix B of \cite{bordag02b} we find the following expression
\begin{align}
  \pi& a \zeta'(0) = \zint dyy\frac{d }{d y}\ln[1-y^2\lambda_0^2(y)]\notag \\
&+ 2\sum_{m=1}^\infty m\zint dyy\frac{d}{dy}\left\{\ln[1-m^2y^2\lambda_m^2
(my)+\frac{y^4t^6}{4m^2}]\right\}\notag \\
  &+ \frac{\pi}{32}\left(3\gamma - 4 +3\ln \frac{a}{2}\right)\label{bordagetal}
\end{align}
with $t = 1/\sqrt{1+y^2}$.
Let us call the two integrals in (\ref{bordagetal}) $X_0$ and $X_m$, where the 
latter is the integral inside the sum. After a partial integration and, in the 
case of $X_m$, a substitution $ym=x$, these can be written on the familiar form
\begin{subequations}
\begin{align}
  X_0=&-\zint dx\ln[1-x^2\lambda_0^2(x)]]\\
  X_m=&-\frac1{m}\zint dx \left\{\ln[1-x^2\lambda_m^2(x)]\right.\notag \\ 
  &+\left.\frac{z^4}{4(m^2+x^2)^3}]\right\}.
\end{align}
\end{subequations}

Comparing with (\ref{Cp}) we see that
\begin{align}
   \pi a \zeta'(0)+2C(1)=&\zint \frac{dxx^4}{4(1+x^2)^3} + \frac{\pi}{32}
\left(3\gamma - 4 + 3\ln \frac{a}{2}\right)\notag \\
   =& \frac{\pi}{64}[6\gamma - 5 + 6 \ln \frac{a}{2}]
\end{align}
since $\frac{1}{4}\zint dxx^4/(1+x^2)^3 = 3\pi/64$. We have thus shown the 
correspondence analytically.

\end{document}